# Three-dimensional non-reciprocal transport in photonic topological heterostructure of arbitrary shape

Mudi Wang[1]*, Ruo-Yang Zhang[1], Chenyu Zhang[2], Haoran Xue[3], Hongwei Jia[1], Jing Hu[1], Dongyang Wang[4], Tianshu Jiang[2], C. T. Chan[1]*

[1]Department of Physics, The Hong Kong University of Science and Technology, Hong Kong, China

[2]MOE Key Laboratory of Advanced Micro-Structured Materials, Shanghai Frontiers Science Center of Digital Optics, Institute of Precision Optical Engineering, and School of Physics Science and Engineering, Tongji University, Shanghai 200092, China

[3]Department of Physics, The Chinese University of Hong Kong, Shatin, Hong Kong SAR, China

[4]Optoelectronics Research Centre, University of Southampton, Southampton SO17 1BJ, United Kingdom

*Correspondence to: mudiwang@ust.hk; phchan@ust.hk

**Electromagnetic wave propagation in three-dimensional space typically suffers omnidirectional scattering when encountering obstacles. In this study, we employed Chern vectors to construct a topological heterostructure, where large-volume non-reciprocal topological transport in three-dimension is achieved. The shape of the cross-section in the heterostructure can be arbitrary designed, and we experimentally observed the distinctive cross-shaped field pattern transport, non-reciprocal energy harvesting, and most importantly, the remarkable ability of electromagnetic wave to traverse obstacles and abrupt structure changes without encountering reflections in 3D space.**

Topological materials have garnered significant attention from researchers due to their remarkable properties, in particular the unidirectional boundary transport [1-15], which makes them highly promising for applications such as optical communications, energy transport, and quantum computing. In comparison to two-dimensional systems, 3D topological materials offer greater degrees of freedom to control the propagation of waves, leading to a wide range of fascinating phenomena such as topological chiral Fermi arcs [16-19], negative refraction [20], Dirac-like surface states [21], antichiral surface states [22], 3D Chern insulators [23], and chiral Landau levels [24,25]. While the topological states are usually robust against defects and disorders, the existing methods only offer unidirectional transport in 1D edge or 2D surface of a topological material. The 3D non-reciprocal bulk transport is highly desired, but remains elusive, both experimentally and theoretically. In this paper, we propose an approach to realize highly robust 3D non-reciprocal bulk transport by designing topological heterostructures [26-30] of arbitrary shape. The embedded non-trivial Chern vectors ensure the immunity to obstacles and sharp boundary shapes, enabling efficient and unidirectional electromagnetic wave propagation in three-dimensional space.

We consider a unit cell as shown in Fig. 1A, which consisted of the yttrium iron garnet (YIG) rod with the height $h_1 = 2\ mm$ and radius $R$, arranged on a metal plate with thickness $h_2 = 2\ mm$ in a hexagonal lattice array. The interlayer coupling strength is controlled by the circular holes with the radii of $r = 2\ mm$. The lattice constant in the *x-y* plane is $a = 16\ mm$, and the periodicity along *z* is $h = 4\ mm$. The 3D structure exhibits "AA" stacking along the *z*-axis, and the remaining space is filled with air. Figure 1B displays the corresponding first Brillouin zone. Crystal A and C are subjected to a magnetic field of 0.1T in opposite directions along the *z*-axis, achieved by embedding the magnets beneath the YIG rods ($R = 2\ mm$) within the metal plates. These magnetic crystals display the characteristics of 3D Chern insulators, exhibiting complete energy band gaps (the upper and lower panels of Figure 1C). In contrast, the YIG rods ($R = 1.65\ mm$) in Crystal B remain unmagnetized and the crystal demonstrates properties of a nodal line semimetal, as depicted in the middle pane of Figure 1C. By combining Crystals A, B, and C in a heterostructure (Figure 1D), which exhibits periodicity in the *x* and *z* directions, we observe the emergence of two sheets of distinct 3D waveguiding topological states (Figure 1E), with the field distribution of the eigenstates spanning the entire 3D region of domain B. Within the frequency range of TNWSs (from 11.7 GHz to 12.6 GHz), the slopes in the $k_x$ direction at both valleys are positive for each fixed value of $k_z$, as shown in Fig. 1F, which exhibits one way transport effects.

Before going further, we should develop the model Hamiltonians to understand the emergence of non-reciprocal waveguide modes in 3D space. The effective Hamiltonians describing domains A, B, and C around the valley near the Dirac points along the high-symmetry line K-L can be given as $H = k_x \sigma_x v_D - i\partial_y \sigma_y v_D + d_z^j \sigma_z$, where $v_D$ represents the slope along both $k_x$ and $k_y$ directions, $d_z^j (j = A, B, C)$ is a function of $k_z$ to characterize the effective mass, and $\sigma_i (i = x, y, z)$ denotes the Pauli matrices. The Hamiltonian form reflects the fact that the system is periodic along *x* and non-periodic along the *y* direction ($k_y$ is not a good quantum number). The band dispersion near the K-L valley is determined by the eigenvalue equation

$$H\psi = \delta\omega\psi. \qquad (1)$$

Domains A and C are 3D Chern insulators, characterized by the vectors of three Chern numbers $\boldsymbol{C}_A = (C_x^A, C_y^A, C_z^A)$ and $\boldsymbol{C}_C = (C_x^C, C_y^C, C_z^C)$, i.e., the Chern vectors [23]. The model Hamiltonian gives $C_x^A = C_y^A = C_x^C = C_y^C = 0$, $C_z^A = sgn(d_z^A) = -1$ and $C_z^C = sgn(d_z^C) = 1$. On the other hand, domain B carries the band dispersion of a typical nodal line semiconductor, which implies $d_z^B = 0$. In the heterostructure ABC, with domain B having a thickness denoted as *L* along the y-direction, the A-B interface is positioned at $y = L/2$, while the B-C interface is located at $y = -L/2$. For frequencies inside the band gaps of domains A (C), the non-reciprocal waveguide states in the heterostructure ABC are expected to exponentially attenuate along the positive (negative) y-direction. The states can be described as $\psi_A = \alpha_A \phi_{k_A} e^{-u_{kA}\left(y-\frac{L}{2}\right)} e^{i(k_x x + k_z z)}$ and $\psi_C = \alpha_C \phi_{k_C} e^{u_{kC}\left(y+\frac{L}{2}\right)} e^{i(k_x x + k_z z)}$. In domain B $(-L/2 \leq y \leq L/2)$, the wave function can be expressed as the superposition of the two linearly independent bulk eigenstates at each $k_x$: $\psi_B = (\alpha_{B1} \phi_{k_{B1}} e^{-u_{kB}\left(y-\frac{L}{2}\right)} + \alpha_{B2} \phi_{k_B} e^{u_{kB}\left(y+\frac{L}{2}\right)}) e^{i(k_x x + k_z z)}$. The coefficients $\alpha_A, \alpha_{B1}, \alpha_{B2}$ and $\alpha_C$ determine the amplitude of each superposed mode. By solving equation (1), we can obtain

$$u_{k_j} = \frac{1}{v_D}\sqrt{k_x^2 v_D^2 - \delta\omega^2 + \left(d_z^j\right)^2}, \quad \phi_{k_A} = \begin{pmatrix} (k_x + u_{kA})v_D \\ \delta\omega - d_z^A \end{pmatrix}, \quad \phi_{k_{B1}} = \begin{pmatrix} (k_x + u_{kB})v_D \\ \delta\omega - d_z^B \end{pmatrix},$$

$$\phi_{k_{B2}} = \begin{pmatrix} (k_x - u_{kB})v_D \\ \delta\omega - d_z^B \end{pmatrix}, \quad \text{and} \quad \phi_{k_C} = \begin{pmatrix} (k_x - u_{kC})v_D \\ \delta\omega - d_z^C \end{pmatrix}.$$ Therefore, the non-reciprocal waveguide mode can be described as:

$$\psi_{ABC} = e^{i(k_x x + k_z z)} \begin{cases} \alpha_A \phi_{k_A} e^{-u_{kA}\left(y-\frac{L}{2}\right)}, & y > \frac{L}{2} \\ \alpha_{B1} \phi_{k_{B1}} e^{-u_{kB}\left(y-\frac{L}{2}\right)} + \alpha_{B2} \phi_{k_{B2}} e^{u_{kB}\left(y+\frac{L}{2}\right)}, & -\frac{L}{2} \leq y \leq \frac{L}{2} \\ \alpha_C \phi_{k_C} e^{u_{kC}\left(y+\frac{L}{2}\right)}, & y < -\frac{L}{2} \end{cases} \quad (2)$$

By considering the continuity of equation (2) at $y = -L/2$ and $y = L/2$, we can get the dispersion of the TNWS: $\delta\omega = k_x v_D$, while $\phi_{k_A}, \phi_{k_{B1}}, \phi_{k_{B2}}$ and $\phi_{k_C}$ are proportional to $\begin{pmatrix} 1 \\ -1 \end{pmatrix}$. Another TNWS around K'-L' valley can be obtained through a similar analysis. Furthermore, it is essential for the transmission direction of the waveguide modes at K'-L' valley to align with those at the K-L valley, as dictated by the Chern number difference $\Delta C_z$ of 2 between domains A and C. These TNWSs can be understood as a combination of the interface state at the A|C domain wall and the bulk states within domain B, resulting in bulk states with non-reciprocal properties.

Now we proceed to the demonstration of 3D non-reciprocal transport. Figure 2A presents a photograph capturing a linear heterostructure, with copper-plated boards positioned at the bottom (Fig. S1A) and top (Fig. S1B). The top board features circular holes for the convenience of field measurements. The numerically computed (using COMSOL) iso-frequency contours, represented by the green dashed line in Figure 2B, appear as two nearly straight lines. To experimentally validate our findings, we conducted measurements on the *xz*-plane within the domain B and performed 2D Fourier transform, generating the Fourier spectrum (the color map in Figure 2B) in the momentum space. Figure 2C illustrates the field distribution in the heterostructure waveguide excited by a vertical line source under varying values of $k_z$, with periodicity observed in the *z* direction. Notably, the energy exhibits a uniform distribution across the entire domain B, indicating remarkable similarity in the field distribution as $k_z$ varies. To evaluate the stability of the observed topological state, we introduced 20 aluminum cylinders as obstacles within the central region (the sample's photo is shown in Fig. S1C), as depicted on the right side of Figure 2C. We observed that the incident wave effectively circumvented these obstacles, allowing propagation with minimal disruption to the field distribution. We measured the transmission characteristics both in the absence and presence of 20 aluminum cylinders, as illustrated in Figure 2D. The obtained results confirm the non-reciprocal nature of transport and its resilience to the presence of obstacles within the frequency range of TNWSs.

The existence of TNWSs in 3D space opens up possibilities for controlling the bulk transport of electromagnetic waves by manipulating the shape of the heterostructure in different directions. We first focus on modifying the shape within the *yz* plane, which is perpendicular to the direction of wave propagation. Figure 3A provides a schematic representation of a cross-shaped heterostructure (Figure S1D shows the corresponding experimental sample). The heterostructure maintains periodicity along the *x* direction, while domain B takes on a cross shape in the yz plane. The projection of the band structure along the x-direction for this cross-shaped heterostructure is illustrated by the green dashed lines in Figure 3B. We should note that, there are a total of 15 distinct one-way bands in each valley, because there are 15 layers in the z-direction. The bands within each valley are so closely spaced that they collectively appear as a single green line in Figure 3B. This phenomenon can be understood using the model Hamiltonian analysis presented in Supplementary Materials.

In the experiment, we conducted measurements along the x-axis and applied a one-dimensional FFT to transform the collected data to momentum space. The resulting plot, depicted in the color map of Figure 3B, demonstrates agreement with the corresponding theoretical predictions. The experimentally measured field distribution in the yz plane is presented in Fig. 3C, reveals an energy distribution taking the form of a cross shape. In comparison to the direct jointing of two different 3D Chern insulators (of opposite chern vectors), the non-reciprocal waveguide states in the presented design can exhibit an arbitrary two-dimensional field distribution perpendicular to the propagation direction as determined by the shape of domain B. More details can be found in Figure S2-5, and the corresponding supplementary text.

We proceed to modify the shape of the heterostructure in the xy plane (being periodic in the z direction), aiming to harvest input energy into a narrow region. Fig. 3D provides a schematic illustration of the heterostructure configuration designed for this purpose. The corresponding experimental sample photo can be found in Fig. S1E. In this particular configuration, domain B within the $xy$ plane abruptly change from a wide area (consisting of 7 layers in the middle) to a narrow area (with 0 layer in the middle, where domains A and C are in direct contact). Examining the simulated results after point source excitation at a frequency of 12.2GHz with different $k_z$, Fig. 3E displays the field distribution where the energy concentrates from a larger area to a smaller area, effectively achieving the line focusing. In our experimental setup, we scanned the $yz$ planes as indicated by the black dashed lines in Fig. 3D. Subsequently, we plotted the distribution of field intensity along the transverse $y$-axis in Fig. 3F, where the contributions of different layers along the z-direction were summed together. As anticipated, the experimental results indicate that during the transmission of microwaves along the *x*-direction, the input energy can be effectively collected to the central region of *y*-direction.

Lastly, we broke the periodicity of the sample in all $x$, $y$, and $z$ directions. In this configuration, the domain B undergoes both bending and deformation, as illustrated in Figure 3G, with the experimental sample photo displayed in Fig. S1F. The entire configuration retains its non-reciprocal bulk transport characteristics, since the topological properties of domains A, B, and C remain unaffected by the changes in configuration. On the other hand, the entire sample can also be viewed as multiple 2D sections being stacked along the *x* direction. As each individual section of heterostructure exhibits non-reciprocal effects, this property will remain valid for the whole structure.

In the experiment, we performed separate measurements of the transmission characteristics for a cross-shaped heterostructure (Fig. 3A), a focusing configuration (Fig. 3D), and a 90-degree curve configuration under deformation (Fig. 3G). The results, depicted in Fig. 3H, reveal a distinct non-reciprocity within the TNWSs region (highlighted by the blue shaded area). This observation demonstrates that despite variations in geometry, the non-reciprocal transport of electromagnetic waves in three-dimensional space remains unaffected.

Our study demonstrates the non-reciprocal transport in 3D photonic heterostructure of arbitrary shape. This is achieved by incorporating a gauge field generated by two different 3D Chern insulators on both sides of a nodal line semimetal crystal. Consequently, the initially reciprocal nodal line semimetal states undergo a transformative change, leading to the emergence of TNWSs. This discovery establishes a platform for the effective control and manipulation of electromagnetic wave transport in 3D space.

## Methods

### Numerical simulation

All theoretical simulations were performed using the RF module of COMSOL Multiphysics. In the calculation of the single-cell band structure shown in Fig. 1C, all directions were set as periodic. For the band calculations in Figs. 1E, F and Fig. S2, the *x* and *z* directions were set as periodic, while the remaining directions were set as PEC boundaries, removing the boundary states at the interfaces between domain A(C) and the PEC boundaries. In the calculation of the projected band structure in Fig. 3B, the *x* direction was set as periodic, while the remaining directions were set as PEC boundaries, removing the boundary states at the interfaces between domain A (C) and the PEC boundaries.

### The calculation of the Chern vector

Since the system is globally gapped, the Chern numbers for the $\hat{x}$, $\hat{y}$ and $\hat{z}$ momentum planes can be well defined by $C_{x,y,z} = \frac{1}{2\pi}\int_{S^{x,y,z}} \Omega_n \, d^2k$, where $S^{x,y,z}$ represents the $k_{x,y,z}$-fixed two-dimensional (2D) section in the 3D BZ. Here $\Omega_n(\boldsymbol{k}) = \nabla_{\mathbf{k}} \times A_n(\boldsymbol{k})$ is the Berry curvature, $A_n(\boldsymbol{k}) = i\langle u_{n,k}|\nabla_k|u_{n,k}\rangle$ is the Berry connection and $u_{n,k}$ is the periodic part of Bloch state for the $n$th band.

### Materials and experimental setups

The YIG material parameters employs a saturation magnetization of $4\pi M_s = 1850\mathrm{G}$, a linewidth of $\Delta H = 50\mathrm{Oe}$, and a permittivity of $\varepsilon = 13.8$. A fully magnetized ferrite has a relative permeability tensor in the form $\bar{\mu} = $

$$\begin{pmatrix} \mu & i\kappa & 0 \\ -i\kappa & \mu & 0 \\ 0 & 0 & 1 \end{pmatrix}, \text{ where } \mu = 1 + \frac{\omega_m(\omega_0+i\alpha\omega)}{(\omega_0+i\alpha\omega)^2-\omega^2} \text{ and } \kappa = \frac{\omega_m\omega}{(\omega_0+i\alpha\omega)^2-\omega^2}.$$ The resonance frequency is denoted as $\omega_0 = \gamma H_{0i}$, the characteristic frequency as $\omega_m = 4\pi\gamma M_s$, and the gyromagnetic ratio as $\gamma = 2.8\text{MHz/Oe}$.

On the copper-plated board, circular holes with a diameter of 3.2mm are drilled at the positions of the YIG. Inside these holes, magnets with a diameter of 3mm and a height of 2mm are placed to introduce the corresponding magnetic field in the desired direction. The YIG is held in place using thin film boards with drilled holes (with a relative permittivity close to 1). The straight heterostructures, cross-shaped heterostructures, and line-focused heterostructures are composed of 26*21*15 unit cells in the *x*, *y*, and *z* directions. While the 90-degree curve configuration while undergoing deformation consists of 26*25*15 unit cells along the x, *y*, and *z* directions.

In the experimental setup, a vector network analyzer is utilized to connect both the source and probe antennas. By inserting the probe into the holes and scanning along the *z* direction, the field information inside the crystal is measured.

## Contributions

M.W. conceived the original idea, performed the simulations and designed the samples. C.T.C. supervised the whole project. M.W., C.Z. and D.W. carried out the experiments. M. W. and R.-Y.Z. performed the analytical derivation. M.W., R.-Y.Z. H. X, H.J, J.H, D. W., T. J. and C.T.C. did the theoretical analysis. M.W., R.-Y.Z. and C.T.C. wrote the draft. All authors discussed the results and contributed to the manuscript.

## Acknowledgements

This work is supported by RGC Hong Kong through grants 16310420, 16307621 and AoE/P-502/20. D.W. acknowledges support from the

Anniversary Fellowship of the University of Southampton. T.J. acknowledges support from the National Natural Science Foundation of China (12304345).

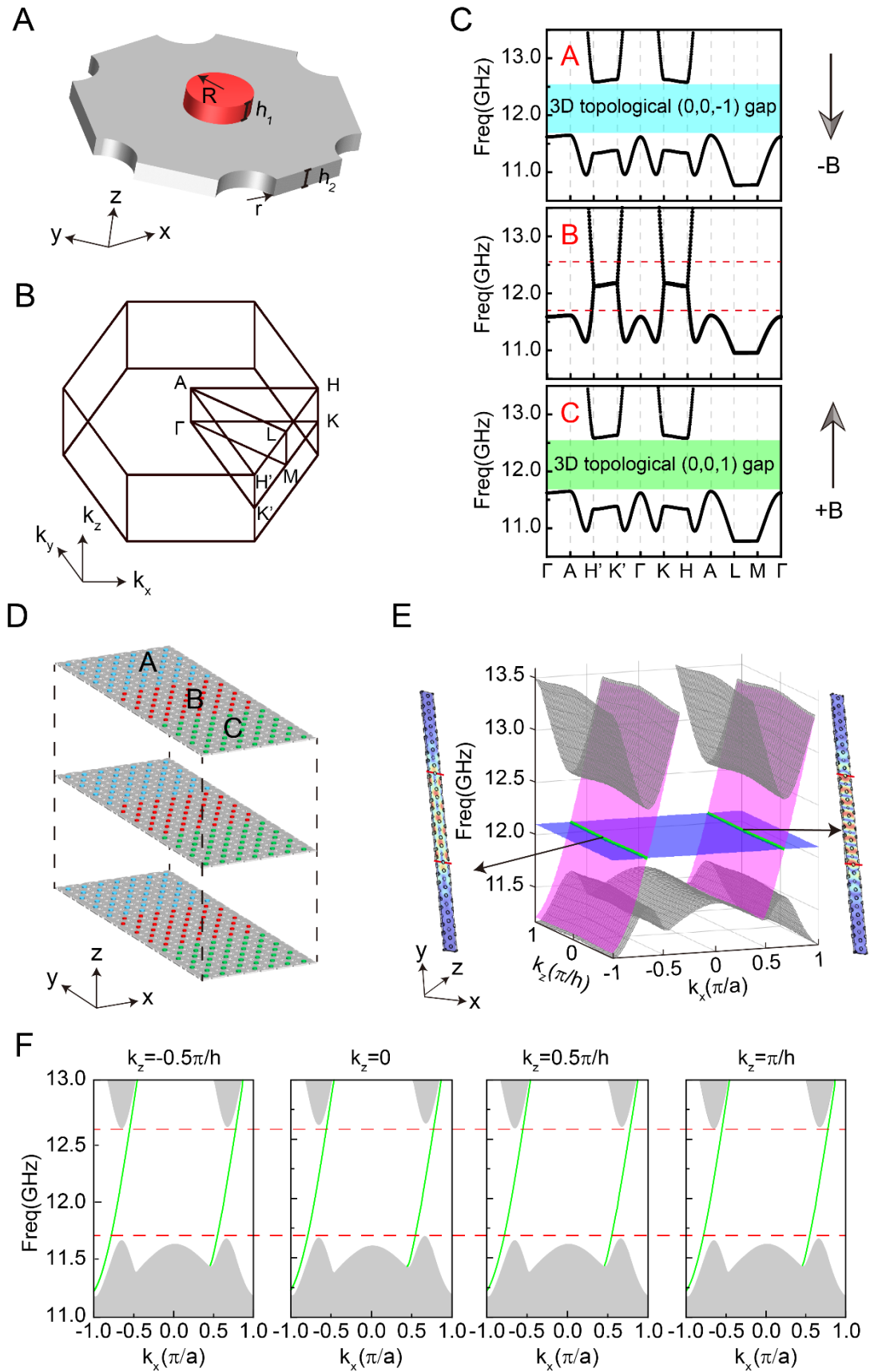

**Fig. 1. Heterostructures for 3D TNWSs. A,** The unit cell of a gyromagnetic

photonic crystal consisted of the YIG rod (red cylinder) and perforated metal plate (gray board). **B,** Bulk Brillouin Zone (BZ). **C,** Band diagrams of the photonic crystal for Crystals A, B, and C. In Crystal A (C), the radius of YIG is $r = 2mm$, and it is subjected to a negative (positive) magnetic field $B = 0.1T$ ($B = -0.1T$). While in Crystal B, the radius of YIG is $R = 1.65mm$, and there is no external magnetic field, hence $B = 0T$. **D,** The schematic diagram of the 3D heterostructures consists of Crystals A, B, and C. **E,** Band diagrams of the heterostructures. A frequency slice was plotted at 12.1 GHz, showing two intersecting green lines with the TNWSs. The illustration demonstrates two eigenstates at $k_z = 0$. **F,** The band dispersion of heterostructure ABC along the $k_x$ direction at different $k_z$ values.

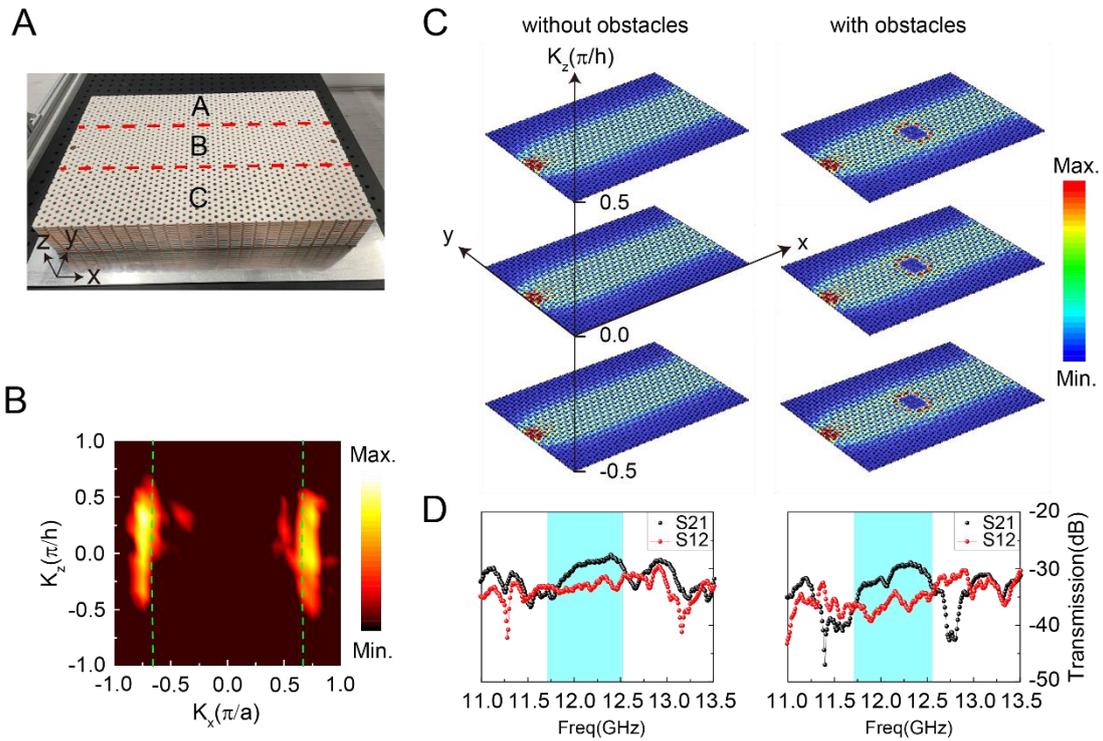

**Fig. 2. Robust transport for 3D TNWSs. A,** Photograph of the straight heterostructure. **B,** Experimentally measured band structure (color map) compared with simulated result (green dash lines) at the $k_x k_z$ plane for 12.1 GHz. **C,** The distribution of field with different $k_z$ at 12.1 GHz for the normal structure and the structure with obstacles (20 aluminum cylinders). **D,** Transmission coefficient S21 for the normal structure and the structure with obstacles.

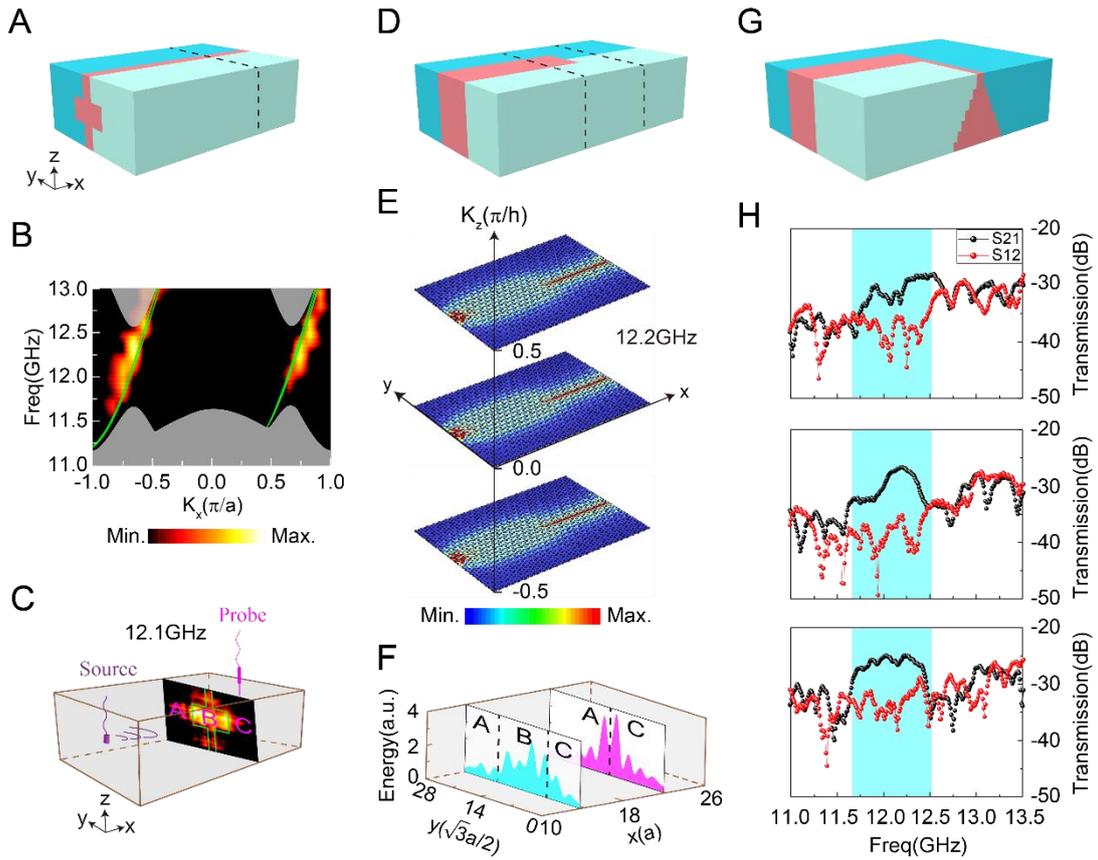

**Fig. 3. Robust transport for 3D TNWSs with arbitrary shape. A,** schematic diagram of a cross-shaped heterostructure. **B,** Experimentally measured projected band structure for the structure in **A**, compared with the theoretical results. **C,** Experimentally measured field distribution for the structure in **A** at 12.1 GHz. **D,** Schematic diagram of a line-focused heterostructure. **E,** Simulated field distribution for the structure in **D** at 12.2 GHz with $k_z$ =-0.5 π/h,0,0.5π/h, with periodicity along the *z*-direction. **F,** Experimentally measured field intensity distribution with different *y* positions at planes X=13a and X=22a for the structure in **D**. **G,** Schematic diagram of 90-degree curve configuration while undergoing deformation. **H,** Transmission coefficients S12 and S21 for the structures in **A**, **D**, and **G**.